\documentclass[superscriptaddress,aps,pre,amsmath,amsymb,reprint]{revtex4-2}
\usepackage{graphicx}
\usepackage{bm}
\usepackage{xcolor}
\usepackage{relsize}
\usepackage{hyperref}
\usepackage{hycolor} 
\hypersetup{breaklinks=true,colorlinks=true,citecolor=[rgb]{0,0.18,0.65}, 
urlcolor=[rgb]{0,0.18,0.65}, linkcolor=[rgb]{0,0.18,0.65}, final = true}

\def\ie{{\it{i.e.,~}}}
\def\kB{{k_{\mathrm{B}}}}
\def\bj{\mathbf{j}}
\def\bx{\mathbf{x}}
\def\by{\mathbf{y}}
\def\T{\mathcal{T}}
\def\xdot{\mathbf{\dot{x}}}
\def\bjst{\bj_{\mathrm{st}}}
\def\pst{p_{\mathrm{st}}}
\def\THETA{\boldsymbol\theta}
\def\Bf{{\mathbf{f}}}

\begin{document}
	\title{Inverse Design of Non-Equilibrium Steady-States: A Large Deviation Approach}
	
	\author{William D. Pi\~{n}eros} 
	\affiliation{Center for Soft and Living Matter, Institute for Basic Science (IBS), Ulsan 44919, Korea}
	\author{Tsvi Tlusty}
	\email{tsvitlusty@gmail.com}
	\affiliation{Center for Soft and Living Matter, Institute for Basic Science (IBS), Ulsan 44919, Korea}
	\affiliation{Department of Physics, Ulsan National Institute of Science and Technology (UNIST), Ulsan 44919, Korea}
	\affiliation{Department of Chemistry, Ulsan National Institute of Science and Technology (UNIST), Ulsan 44919, Korea}

	\date{\today}
	
	\begin{abstract}
	The design of small scale non-equilibrium steady states (NESS) is a challenging, open ended question. 
	While similar equilibrium problems are tractable using standard thermodynamics, a generalized description for non-equilibrium systems is lacking, making the design problem particularly difficult. 
	Here we show we can exploit the large deviation behavior of a Brownian particle and design a variety of geometrically complex steady-state density distributions and flux field flows. 
	We achieve this design target from direct knowledge of the joint large deviation functional for the empirical density and flow, and a ``relaxation'' algorithm on the desired target states via adjustable force field parameters. 
	We validate the method by replicating analytical results, and demonstrate its capacity to yield complex prescribed targets, such as rose-curve or polygonal shapes on the plane.
	We consider this dynamical fluctuation approach a first step towards the design of more complex NESS where general frameworks are otherwise still lacking.  
	
 	\end{abstract} 

	\maketitle 

\section{Introduction} 
The ability to create and \emph{design} a system to a desirable goal stands as the culmination of deliberate human control and precise, applied knowledge. 
While this level of achievement is generally possible at comparative macro scales where bulk properties dominate, similar control is lacking for systems below the micro-scale where fluctuations, thermal or otherwise, is significant. 
Notwithstanding, the obvious appreciation of biological complexity and its myriad of small scale solutions to practical problems~\cite{GeckoFeetStructureForce,MotorProteins_theory}, for instance, suggests the design barrier is not insurmountable. 
Indeed, design problems involving small systems have been an active topic of research in soft or condensed materials~\cite{ID_NanoPhotonicsRev,ID_MaterialRev,ID_general}, 
motivated by the rising synthetic and technological prowess at this scale~\cite{2DNanoFabricationRev,PhotonicMatsDesign,SelfAssemblyPerspective,SelfAssemblyReviewColloids2}.

Broadly, one can envision the design problem as a generic optimization goal where one seeks to fulfill an objective subject to constraints that reflect the desired state of a system. 
Solution follows through standard numerical optimizers where possible. Such formulations can also be heuristic~\cite{HeuristicOptRev,HeuresticOpt_ConfProb}, or inferential, e.g. learning algorithms~\cite{ML_broadmolecularRev, ML_softmatterRev, ML_softmatterRev2}, and are practical in a variety of systems in so far as they can yield an answer~\cite{ToyModelProteinFoldingInverseDesignHeuristic,InorganicMatInvDesignML,QuatumDotSpectraInvDesign}. 
However, in framing the design problem, it is desirable to maintain a clear physical picture so as to better understand the nature of the solution and its optimization process.  
For instance, in the equilibrium case, one usually seeks to minimize a free energy of a system, for example by modifying the underlying interactions~\cite{InvDesignTechRev,ID_SoftMatterRev}.  Indeed, such an approach can be applied successfully to the design of a variety of self-assembling phases~\cite{template_directed_free_energy,DiamondGroundStateID,Truncs_gs_opt,GrandCanonicalCrystalDesign}.

Extending a similar approach to the problem of design of non-equilibrium systems at the small scale is a critical yet mostly unaddressed issue.
	While a fundamental framework akin equilibrium thermodynamics remains elusive, some general statements at the dynamical fluctuations level such as fluctuation theorems are known~\cite{FluctuationTheoremsStochasticDynms,FluctuationTheoremsRev,FluctuationTheoremsMathRev,StochFluctuationTheoremRev,StochasticThermodynamicsRev}, and experimentally demonstrated~\cite{RNACrooksRelation,SingleColloidParticleExp,ExperimentalFluctuationTheoremSS}.
	In particular, to make a non-equilibrium problem more tractable, one generally restricts attention to steady-states, which are time invariant and therefore amenable to averaging. 
	Critically, what distinguishes a non-equilibrium steady state (NESS) from an equilibrium one is the presence of non-vanishing fluxes. 
	Therefore, in addition to the usual probability measure of a system's configuration, a viable statistical description must also account for the system's fluxes. 

	In this regard, the theory of large deviations can fill the gap and provide a window into a NESS system from a dynamical level. 
	Chiefly, the large deviations framework concerns with the deviation of a quantity from its most likely expectation in the limit of infinite sampling. 
	Specifically, a stochastic quantity $x$ fulfills a large deviation principle if its probability measure scales as $P(x) \approx \exp(-n I(x))$ in the large sample limit 
	\footnote{Formally, the limit of $-\frac{1}{n} \ln[P(x)$] as $n\to\infty$ defines $I(x)$.}, 
	where $n$ is the number of samples, and $I(x)$ is a so-called rate function~\cite{LargeDevCurrentDensity,FormalLargeDev2.5}. The most likely value of $x$ is therefore the minimum of the rate function. 

	Importantly, rate functions provide a compact statistical representation of a general stochastic process, where information about the average and variances are readily available.
	Indeed, large deviation principles are applicable for equilibrium systems in the large particle limit where rate functions enter in the form of ensemble specific expressions. For instance, in the canonical ensemble the rate function takes the role of a free energy, and the equilibrium state is recovered as a constrained minimum of this rate function (the most likely state)~\cite{LargeDevStatMech}.
	Thus, knowledge of a rate function can provide a practical statistical description of a system's asymptotic behavior without the detailed tracking of a complex dynamics. For a rigorous definition of rate functions and technical treatment of large deviations, we refer to existing literature~\cite{LargeDevBook,LargeDevGeneralRev,LargeDevRev2}.

	Here we concern ourselves with a NESS dynamics for which a large deviation principle exists, such that information about its configurations \emph{and flux} are available via a rate function. 
	In particular, we consider the paradigmatic case of a colloidal particle in a viscous bath whose dynamics can be modeled through an overdamped Langevin equation:
\begin{equation}
	\dot{\bx} = \chi \mathbf{f(x)} + \left( 2 \chi \beta \right)^{-1/2} \boldsymbol\xi(t)~,
\end{equation}
where $\dot{\bx}$ is the velocity vector of a particle, $\bx$ its position, $\mathbf{f(x)}$ are forces, $\chi$ is the mobility, $\beta=1/\kB T$ is the inverse bath temperature ($\kB$ set to unity hereafter), and $\xi$ is a standard white noise term. Non-equilibrium is achieved by means of non-conservative forces applied on the particle. 

For these type of Markovian diffusion systems, the fluctuations can be captured as a joint probability of configurations and fluxes~\cite{RateFuncDerivation1,RateFuncDerivation2}, 
in the form of a rate functional:
\begin{equation} 
	I[\rho,\bj] = 
	\frac{\beta}{4} \int d\bx \rho(\bx)\left[
	\frac{\bj(\bx)}{\rho(\bx)} - 
	\frac{\bj_{\rho}(\bx)}{\rho(\bx)} 
	\right]^2~,
\label{eq:ratefunc}
\end{equation} 
where  $\bj_{\rho}(\bx)$ is given by 
\begin{equation}
 	\frac{\bj_{\rho}(\bx)}{\rho(\bx)} = 
	\frac{\bjst(\bx)}{\pst(\bx)} 
       -\frac{1}{\beta}\nabla \ln \frac{\rho(\bx)}{\pst(\bx)}~.
\label{eq:jrho}
\end{equation} 
In Eqs.~(\ref{eq:ratefunc}-\ref{eq:jrho}), $\pst$ is the most probable steady-state occupation probability, and the function $\rho(\bx)$ is an empirical density of spatial configurations defined as the time average over a trajectory $\bx(t)$ during time period $\T$ as 
\begin{equation}
	\rho(\by;\bx) = \frac{1}{\T}\int^{\T}_0{ dt \; \delta(\by-\bx)}~.
\end{equation}
Similarly, $\bj_{\mathrm{st}}(\bx)$ is the steady state flux and $\bj(\bx)$ represents the empirical flux or current defined as 
\begin{equation}
	\bj(\by;\bx) = \frac{1}{\T}\int^{\T}_0 dt \; \xdot(t) \delta(\by-\bx)~,
\end{equation} 
where the integral is taken in the Stratonovich sense. 
In short, these empirical measures are finite-time averages of the system, where the true densities are achieved in the long time limit.
Altogether, the rate functional $I$ in Eq.~(\ref{eq:ratefunc}) reflects the most relevant information on the expected steady-state system, in terms of its configuration density and flux. 

In this work, we exploit knowledge of the Brownian particle's rate functional, 
and show we can leverage the dynamical fluctuations to design non-equilibrium steady states. 
Specifically, we imagine a target state as a far-off fluctuation that -- 
through adjustment of appropriate force-field parameters $\THETA$ -- 
``relaxes" as the most probable steady-state,
i.e. minimizes the rate function. 
We carry out this parameter update directly from a Brownian simulation,
given a prescribed form of the configuration and flux configuration that represents a steady-state target. 
We remark that this approach differs from numerous force field fitting studies of Brownian systems, where the goal is to replicate known trajectories or experimental data, and are usually done at a statistical inference level~\cite{FFinf_Experimental,FFinf_MaxLhood1,FFinf_MaxLhood2,FFinf_Bayes,FFinf_Bayes2,FFinf_BrownianSignal}.
Instead, here we focus on \emph{creating} arbitrary NESS in the form of specific configuration probabilities and fluxes, which we do through direct knowledge of the system's dynamic behavior. 
As we will elaborate below, this method can recover known analytical answers and help directly realize non-trivial steady-states.

The rest of the paper is structured as follows. 
We elaborate on the mathematical and simulation details of our method in Sec. \ref{sec:Methods}. 
In Sec. \ref{sec:Results} we first validate our method against a simple analytical scenario. We then show we can realize various prescribed NESS featuring complex occupation probabilities and fluxes in the shape of rose-curve and polygon motifs.
We also offer some thoughts on a physical interpretation of the method.
We end with some conclusions and offer some possible future directions in Sec. \ref{sec:Conclusion}.

\section{Methods}
\label{sec:Methods}
\subsection{General Design Approach}
\label{ssec:design}
We approach the non-equilibrium steady-state design problem from a fluctuation dynamics point of view, using the rate functional in Eq.~(\ref{eq:ratefunc}). 
We envision the steady-state target as a far-off fluctuation that is ``relaxed'' via adjustments of a force field parameters $\THETA$ so as to become the most probable steady-state i.e. $\rho(\bx)=\pst(\bx)$ and $\bj(\bx)=\bjst(\bx)$. 
This requires that the rate function be minimized relative to parameters such that $\nabla_{\THETA} I=0$ and therefore $I=0$. 
We demonstrate this goal can be achieved via an iterative gradient descent method on the parameters by minimization of the rate functional as follows.

First we re-write the rate functional in Eq.~(\ref{eq:ratefunc}), using Eq.~(\ref{eq:jrho}), as 
\begin{equation}
\begin{aligned} 
	I[\rho,\bj] &=
		\frac{\beta}{4} \int{ d\bx \rho(\bx)
		\left[\frac{1}{\beta} \nabla \ln{ \frac{\rho(\bx)}{\pst(\bx)}} \right ]^2 } \\
	       &+ \frac{\beta}{4} \int{ d\bx  \rho(\bx)		
		\left[\frac{\bj(\bx)}
		{\rho(\bx)}-\frac{\bjst
		(\bx)}{\pst(\bx)} \right ]^2} \\
			   & \equiv I_A[\rho] + I_B[\rho,\bj]~,
\end{aligned}
\end{equation}
where $I_A$ depends on probability $\rho(\bx)$ alone and $I_B$ contains the flux field dependent terms. 
We identify the fluctuations $\bj(\bx)$ and $\rho(\bx)$ as prescribing the target state, and let $\pst(\bx|\THETA)$ and $\bjst(\bx|\THETA)$ be intermediate steady-states controlled through $\THETA$. 

We can now update parameters $\THETA$ via a standard iterative gradient descent procedure as 
\begin{equation}
\THETA^{i+1} = \THETA^{i} 
- \alpha_A \nabla_{\THETA^i}I_A(\THETA^i)
- \alpha_B \nabla_{\THETA^i}I_B(\THETA^i)~,
\label{eq:graddescent}
\end{equation} 
where $\THETA^i$ indicates vector of parameters at iteration $i$, and $\alpha_A$ and $\alpha_B$ are the respective gradient update step sizes. 
The parameter gradients themselves can be evaluated as
\begin{equation}
\begin{aligned}
\nabla_{\THETA^i}I_A(\THETA^i) & = \\ 
		  -\frac{1}{2 \beta}  \int d\bx & \rho(\bx) 
		 \left[\nabla \ln \frac{\rho(\bx)}{\pst(\bx| \THETA^i)} \right ]\cdot 
		 \nabla_{\THETA^i}\nabla \ln \pst(\bx| \THETA^i)~,\\
\label{eq:dIA}
\end{aligned}
\end{equation} 
and
\begin{equation}
\begin{aligned}
\nabla_{\THETA^i}I_B(\THETA^i) &  = \\ 
		\frac{\beta}{2} \int d\bx & \rho(\bx) 
		 \left[ \frac{\bj(\bx)}{\rho(\bx)}
                 -\frac{\bjst(\bx| \THETA^i)}{\pst(\bx| \THETA^i)} 
                 \right ]\cdot 
		 \nabla_{\THETA^i }
                 \left ( \frac{\bjst(\bx| \THETA^i)}{\pst(\bx| \THETA^i)} \right )~.\\
\label{eq:dIB}
\end{aligned}
\end{equation} 
The effective steady-state velocity $\bjst(\bx| \THETA^i) / \pst(\bx| \THETA^i)$ in Eq.~(\ref{eq:dIB}) can be computed from the Fokker-Planck expression 
$\bjst/\pst = \Bf - {\beta}^{-1} \nabla \ln \pst$ as 
\begin{equation} 
\nabla_{\THETA^i }\left ( \frac{\bjst(\bx| \THETA^i)}{\pst(\bx| \THETA^i)} \right ) =
\nabla_{\THETA^i }\Bf(\bx| \THETA^i)
-\beta^{-1} \nabla_{\THETA^i }\ln \pst(\bx| \THETA^i)~.
\end{equation} 
Each of the integrals in Eqs.~(\ref{eq:dIA},\ref{eq:dIB}) can be evaluated with knowledge of the fixed target states [$\rho(\bx)$, $\bj(\bx)$], adjustable force field $\Bf(\THETA^i)$, and intermediate steady states $\pst^i$, the latter which are found directly from a Brownian simulation.

The optimization procedure is then carried out in the following fashion: 
(A) given an initial guess of the force field $\Bf(\THETA^i)$ compute $\pst^i$ from a Brownian simulation; 
(B) Use $\pst^i$ to evaluate gradients in Eqs.~(\ref{eq:dIA},\ref{eq:dIB});
(C) update parameters as per Eq.~(\ref{eq:graddescent}); 
(D) repeat iteration using updated $\Bf(\THETA^{i+1})$. 
The optimization then completes when gradients converge to zero.

As a model of an adjustable force field $\Bf$ (parameters left implicit), we consider 
\begin{equation} 
	\Bf = \Bf_h + \Bf_{an} + \Bf_o +\Bf_s ~,
\label{eq:f_general}
\end{equation}
where $\Bf_h$ is a confining harmonic force, $\Bf_{an}$ is a driven angular force, $\Bf_o$ is an isotropic repulsive term at the origin, and $\Bf_s$ are a series of variable sinusoidal terms. 
Thus, the system can be interpreted as a confined particle subject to complex sinusoidal forces which constrain it to a specific occupation shape and flow.  
For further details of the optimization implementation, Brownian simulation and exact force-field choices as well as solutions, see the Supplementary Material. 

\subsection{Design Targets}
\label{ssec:targets}

\begin{table}
\caption{\textbf{Design target construction parameters}. For the rose-motif target, these correspond to $k$, the number of petals, and $a$ - petal sharpness. 
For all roses, $r_0=1.25$ is the petal size and $\sigma^2=0.5$ is the variance of the Gaussian-like function. 
For polygons, $l$ is the side-length size for the respective equilateral polygon centered at the origin, and $\sigma^2$ is the variance of the brush Gaussian profile (see text). }
\label{tab:params}
\begin{ruledtabular} 
\begin{tabular}{lcc}
 	& Roses 		& Polygons 	 \\ \hline
1	&$k=3$;$\;$  $a=0.4$ 	& hexagon; $l=1.25$, $\sigma^2=0.30$ \\
2	&$k=4$;$\;$  $a=0.4$ 	& square;$\;\;\;$   $l=1$,$\quad\;$   $\sigma^2=0.25$ \\
3	&$k=12$; $a=0.3$ 	& triangle;$\;$ $l=1.20$, $\sigma^2=0.45$ \\
\end{tabular}
\end{ruledtabular}
\end{table}

Having defined the design strategy, we now consider concrete design targets that highlight the efficacy of the procedure. 
To this end, we consider a two dimensional colloidal particle such that its average occupation and flow trace a specific figure motif on the plane. 
As a first example, we consider a family of ``rose" curves resembling flower petals: 
\begin{equation} 
r_c(\phi)=r_0 \left( 1-a\cos(k \phi) \right)~, 
\label{eq:rosecurve}
\end{equation}
where $r_c$ is the radial coordinate, $\phi$ is the polar angle, $r_0$ is the size of the petals, $k$ is the number of petals, and $a\in(0,1]$ controls petal sharpness. An illustration is provided in the inset of Fig.~\ref{fig:k4rep}a below. Target probability density is then defined as a normalized Gaussian-like function 
\begin{equation} 
p_{\mathrm{tgt}}(r,\phi) = \frac{1}{N} 
\exp\left[ -\frac{(r-r_c(\phi))^2}{2\sigma^2} \right]~,
\end{equation}
where $N$ is the normalization factor, and $r_c$ is defined by the rose curve in Eq.~(\ref{eq:rosecurve}). 

Defining a target flux field $\bj_\mathrm{tgt}$ is somewhat more challenging since, by definition, $\nabla\cdot \bj_\mathrm{tgt}=0$. 
Thus, creating a vanishing divergence is non-trivial, and instead we approximate the flux field as a flow tangential to the chosen motif figure
\footnote{Of course, exact knowledge of both the density and flux field would define the force field through the Fokker-Planck expression of the flux. However, in such a case, the method's purpose would still lie in finding a simpler, closed-form force model from otherwise fully prescriptive targets.}.  
As we will see, this procedure ensures that the prescribed flux field remains plausible as a realizable target. 
Final chosen target parameters are shown in Table \ref{tab:params}.  

Additionally, we consider a second family of more challenging design targets in the form of polygonal motifs. 
The probability $p_{\mathrm{tgt}}(r,\phi)$ is now piece-wise continuous and drawn as a polygon brush whose size is scaled linearly along the plane,
and its paint intensity is determined by an angular symmetric Gaussian profile  $\exp{[-(s-s_0)^2/2\sigma^2]}$, where $s$ is the scaling and $s_0=1$. 
As in the rose-curve case, the flux field is chosen as a uniform flow tangent to the polygon shape.  
Chosen polygonal targets are equilateral triangle, square, and hexagon with parameters as shown in Table \ref{tab:params}.

\section{Results and Discussion} 
\label{sec:Results} 

\begin{figure*}[!htb]
\includegraphics[scale=0.29]{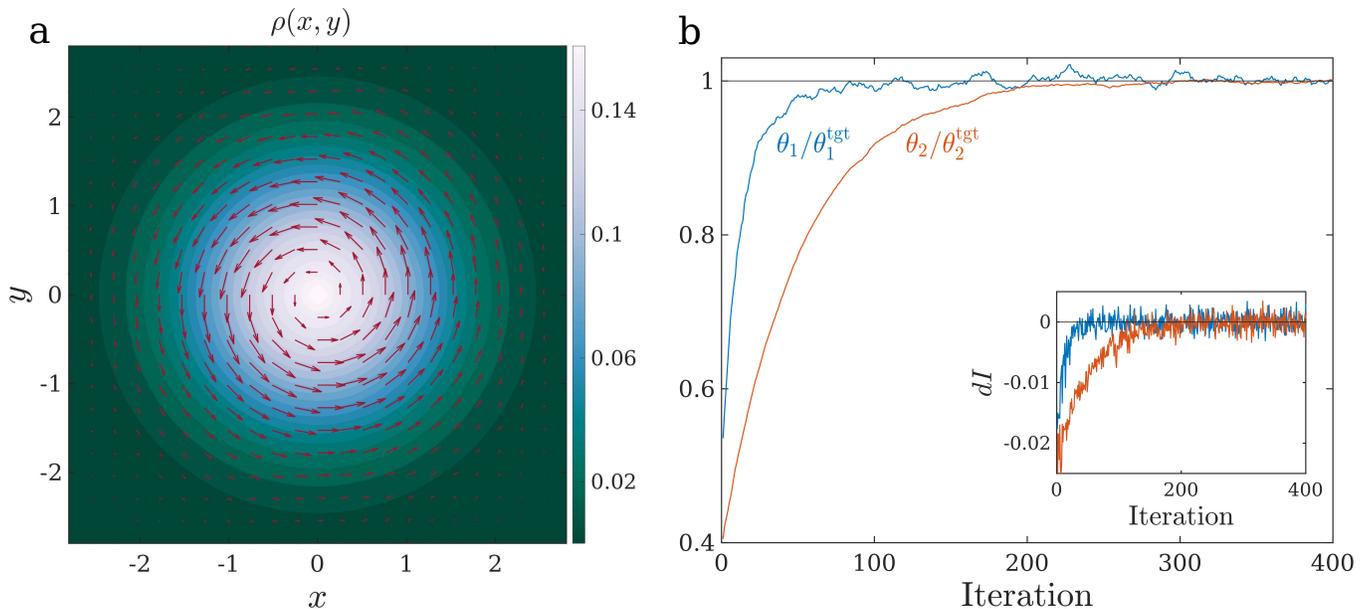} 
\caption{
(a) Test case: a particle in a harmonic potential with a driven angular field whose respective strengths are controlled by parameters $\theta_i$ (Eq.~(\ref{eq:f_harmang})). Analytic solution is given by a Gaussian probability and an angular flux (Eq.~(\ref{eq:test_solutions})).
(b) Progression of the parameters along the iterative gradient descent procedure, for targets defined as the steady state solution of Eq.~(\ref{eq:f_harmang}) with $\beta=2$, $\theta_1^{\mathrm{tgt}}=0.5$, $\theta_2^{\mathrm{tgt}}=2$, where parameters $\theta_i$ are randomly initiated. 
Solution is achieved when $\theta_i/\theta_i^{\mathrm{tgt}}=1$, corresponding to $dI \equiv \nabla_{\THETA}I_{A/B} \to 0$ (inset).}
\label{fig:harmex}
\end{figure*}

We begin by proof-testing the method outlined in Sec. \ref{sec:Methods}.
To this end, we consider the simple analytical case of a particle confined by a radial harmonic potential and driven by an angular force along the plane. 
The force-field is 
	\begin{equation} 
		\Bf =  -\theta_1 r \; \mathbf{\hat{r}} + 
		\theta_2 r \; \boldsymbol{\hat{\phi}}~,
	\label{eq:f_harmang}
	\end{equation}
where the first term is a harmonic force with confining constant strength $\theta_1$, and the second is an angular force with driving magnitude $\theta_2$. 
The analytical steady state solution of this system is known~\cite{RateFuncDerivation1}
(Fig.~\ref{fig:harmex}a) :
\begin{equation}
\pst(r) = \frac{\beta\theta_1}{2\pi} \exp\left[-\frac{\beta\theta_1}{2} r^2 \right] 
\; ,\qquad 
\bjst (r) = \pst(r) \theta_2 r \; \boldsymbol{\hat{\phi}}~,
\label{eq:test_solutions}
\end{equation}
where $\beta$ is the inverse bath temperature. 
Thus, given a chosen set of $\THETA^{\mathrm{tgt}}=\{\theta^{\mathrm{tgt}}_1,\theta^{\mathrm{tgt}}_2\}$, we know the resulting steady state density $\pst$ and $\bjst$ which we can set as test design targets. 
Then, using the model force with tunable parameters $\THETA=\{\theta_1,\theta_2\}$ in Eq.(\ref{eq:f_harmang}),
we can check if the method we propose finds the prescribed $\THETA^{\mathrm{tgt}}$ values.  
The results of this test are shown in Fig.~\ref{fig:harmex}b, where the adjustable $\theta_i$ are initiated randomly and their convergence progress is monitored as a function of iteration number. 
As seen, the ratio of $\theta_i/\theta_i^{\mathrm{tgt}}$ converges to unity past around $300$ iterations, indicating successful parameter solution. 
This convergence corresponds to the vanishing of the rate functional gradients, $dI \equiv \nabla_{\THETA} I_{A/B}$ (inset) indicating rate function minimization. 
	
\begin{figure*}[!htb]
\includegraphics[scale=0.45]{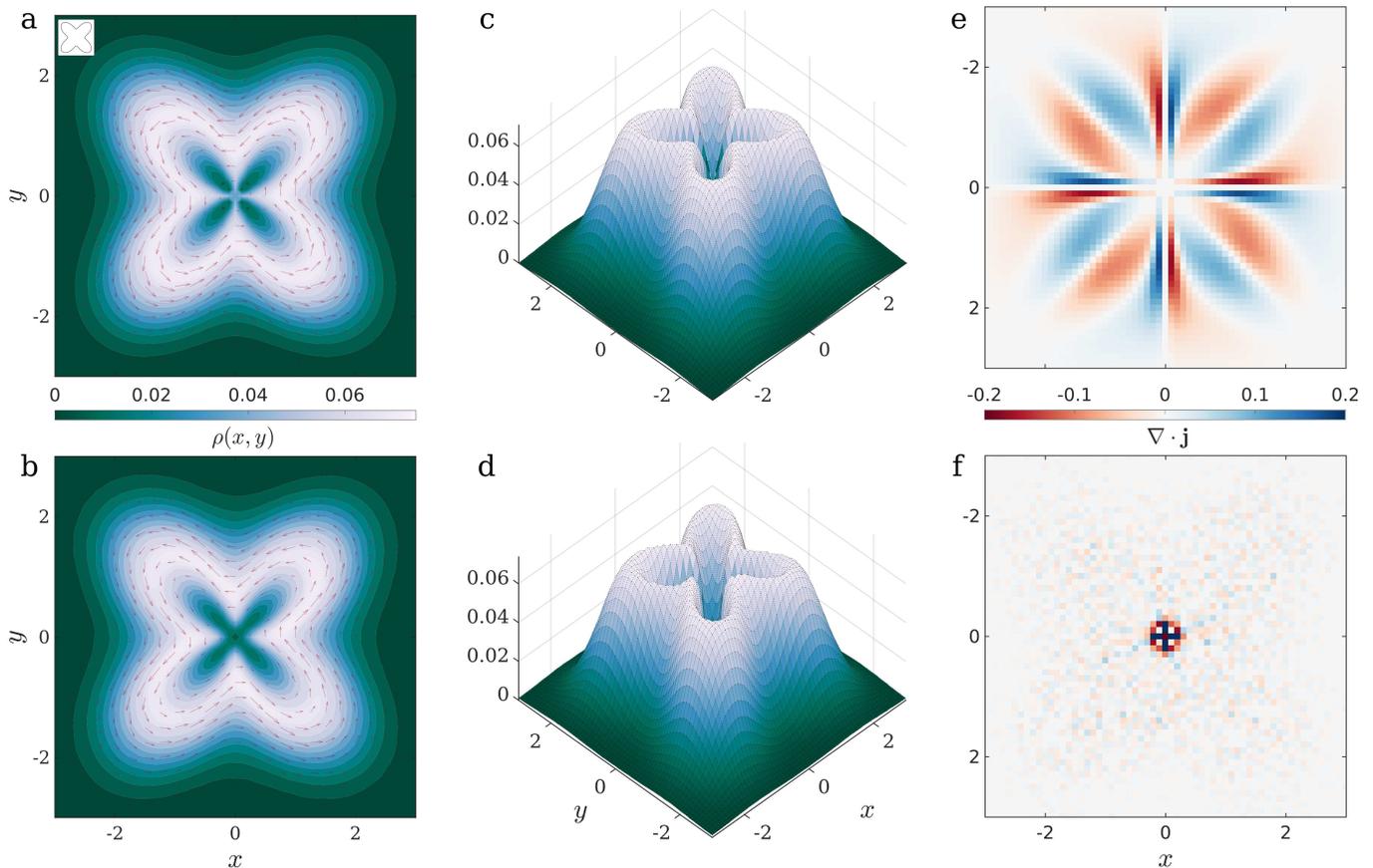} 
\caption{Representative example of a rose-curve target steady state ($k=4$).
(a) Target probability density $\rho(x,y)$ and flux field (red arrows). 
(b) Realized target probability density and flux. 
Small inset is the parametric rose-curve plot (Eq.~(\ref{eq:rosecurve})) with $r_0=1.25$, $k=4$, and $a=0.4$. 
(c-d) Perspective views of (a) and (b) respectively. 
(e) Divergence of the flux $\nabla\cdot \bj$ for the input target and the realized result in (f). 
While the input is divergent, the realized result from a simulation obtains a valid flux field by default. Non-zero values in (f) are a finite grid size artifact.}
\label{fig:k4rep}
\end{figure*} 

Next, 
we move on to consider more complex steady-state design targets. 
To this end, we first start with the rose-motif set of targets which, as elaborated in Sec. \ref{ssec:targets}, represent particle confinement and flow constrained to trace-out a flower-like pattern along the plane. 
As a representative case, we consider the four petal rose curve ($k=4$) whose respective target $\pst$ and $\bjst$ as are illustrated in contour projection
(Fig.~\ref{fig:k4rep}a) and relief representation (Fig.~\ref{fig:k4rep}c). 
As seen in the second figure row in (b and d), our method successfully replicates the prescribed probability and flux field density with excellent agreement. 
Importantly, while the input target field is not entirely valid as seen by the non-zero divergence (Fig.~\ref{fig:k4rep}e), the realized field not only approximates the desired design flow, it also fully realizable, \ie divergence-free (Fig.~\ref{fig:k4rep}f). 
This follows directly from the simulation component of the method, where parameters can only be adjusted with knowledge of the intermediate state probability and flux field densities. 
Thus, the difficulty of creating divergenceless flux-fields as a design target can be averted by ensuring the target flux-field is simply plausible within the motif constraint. 
\begin{figure*}[htbp!]
\includegraphics[scale=0.7]{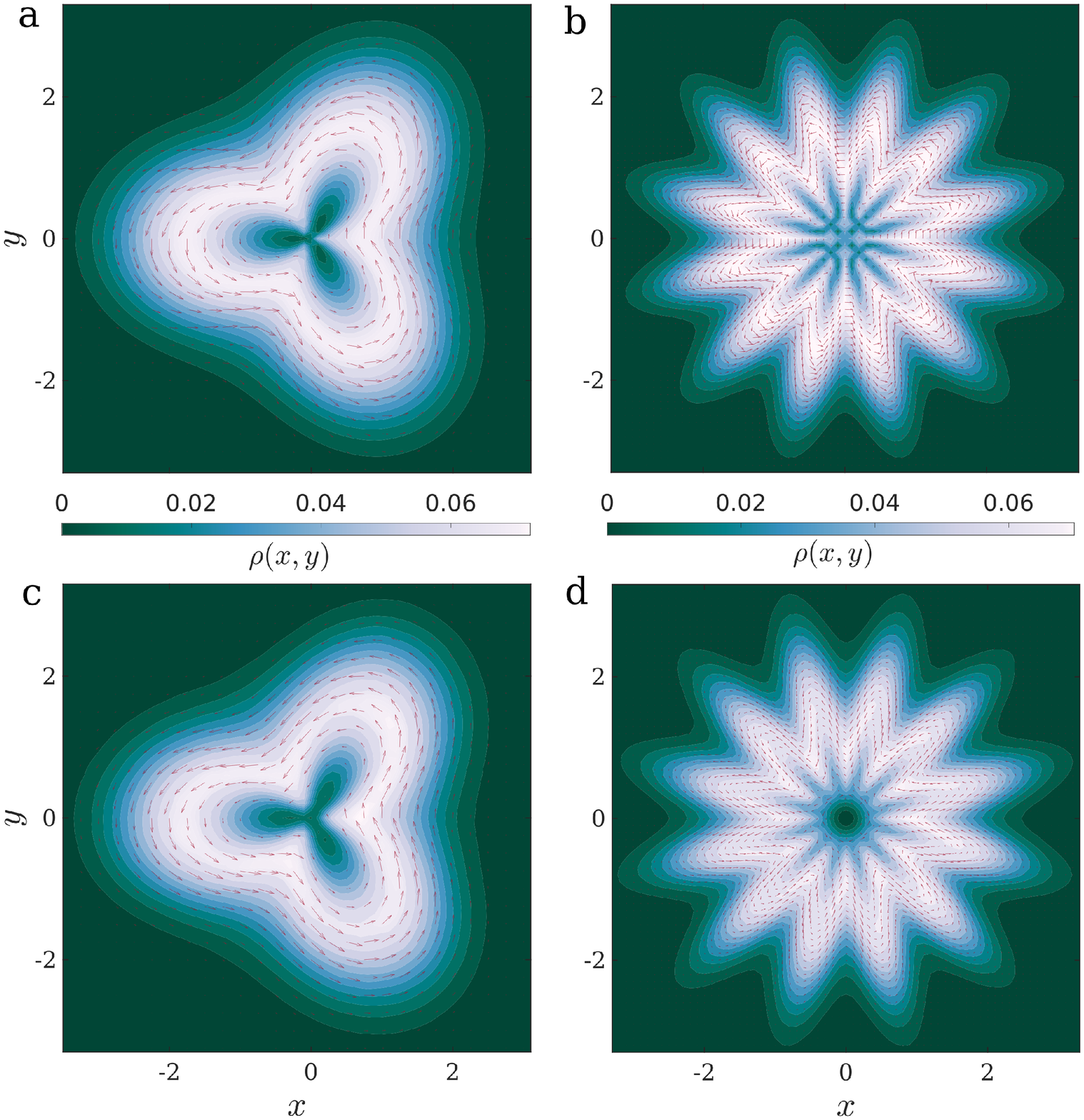} 
\caption{Sample probability $\rho(x,y)$ and flux field targets (red arrows) using the rose-curve motif for (a) $k=3$ and (b) $k=12$. (c) and (d) are the respective realized solutions using our method.
}
\label{fig:roses}
\end{figure*} 

In Fig.~\ref{fig:roses}, we show two additional examples of designed rose-motif targets for an odd-number of petals $k=3$ in (a), and a more complex rose pattern with $k=12$ in (b). 
As seen in (c) and (d), both targets are successfully designed with both flux fields conforming to the imposed motif. 
Notably, in the case of $k=12$, the target displays rapidly changing flux fields with tight changes around every petal turn. 
Despite this, the generated flux field nicely conforms to figure and naturally smooths out the more jarring, and unrealizable, kinks in the original input target. 
\begin{figure*}[!htb]
\includegraphics[scale=0.445]{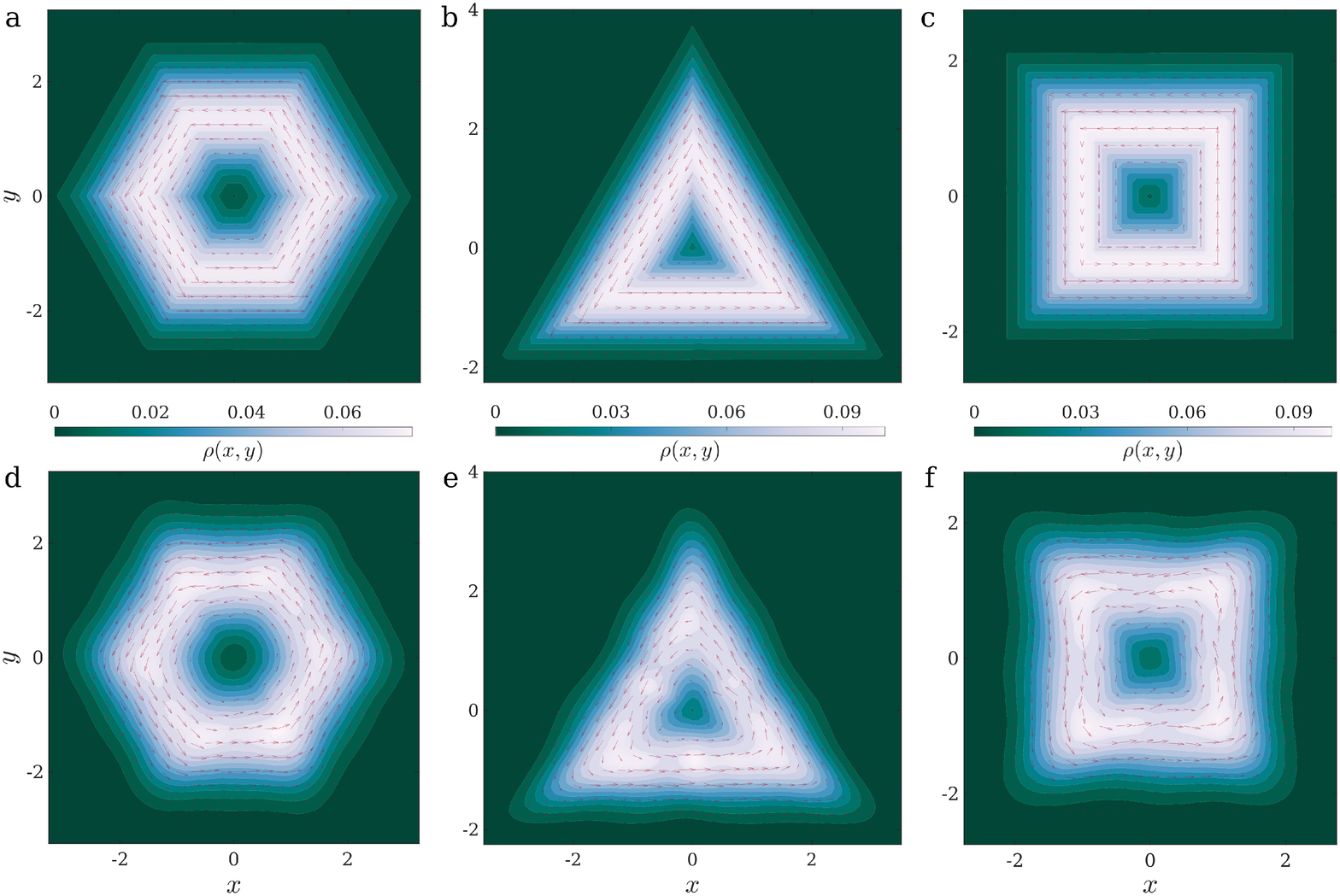} 
\caption{Piece-wise continuous targets with polygonal motifs in position probability density $\rho(x,y)$ and flux fields (red arrows). 
(a-c) input targets for hexagon, triangle and square polygon motifs, respectively. 
(d-f) Realized probability density $\rho(x,y)$ and flux fields (red arrows).}
\label{fig:polygons}
\end{figure*} 

Next, we consider more challenging motifs tracing polygon-shaped densities and field fluxes. 
Unlike the smooth rose-motif targets, here piece-wise continuous, rectilinear edges are imposed on the steady-state targets. 
Due to the singularities, one expects such hard-cut densities and flows to be only asymptotically realizable. 
Yet, assuming a sufficiently flexible force field is provided, we expect the method to find the best approximate realization. 
Indeed, as seen in Fig.~\ref{fig:polygons}, probability densities and flux fields displaying (a) hexagonal, (b) triangular and (c) square  motifs are realized. 
Notably, the difficulty in achieving such targets is observed in the solutions (d-f), where the hard-edged corners become rounded, and the rectilinear flows along the edges exhibit small deviations along the tracks. 
The triangle motif proved particularly difficult to solve, requiring larger sampling times, double phase terms, and additional sinusoidal angular driven terms. 
The difficulty is mostly due to the acute angles that impose stronger singularity and deviation from smoothness, as well as the reflection asymmetry along the x-axis. 

We end with the following tentative thoughts and observations. While our method works well for the design of visually appealing steady-states, its practicality is limited to the explicit knowledge of a rate function of the relevant dynamics, such as interactionless Brownian particles. Extending this approach to the more interesting case of inter-particle interactions relegates the problem to the establishment of an appropriate rate function which can be challenging outside of system approximations like mean-field interactions~\cite{LargeDevGeneralRev,LargeDevMeanField}. 
What is notable, however, is that it is indeed possible to (1) conceive a desirable target state as a dynamical fluctuation and (2) realize it as the most likely steady-state following an appropriate parameter ``relaxation''.

This fluctuation-relaxation picture is, perhaps, not entirely surprising given its central role in the fluctuation-response theory for near-equilibrium processes, where the most likely relaxation pathway following a perturbation also corresponds to an equivalent spontaneous fluctuation pathway of the system~\cite{FluctuationRelaxation1,FluctuationRelaxation2}.  
In fact, it recently came to our attention that the rate function itself can be estimated tightly with a mapping of optimal target states through a variational procedure, consistent with a reciprocal picture ~\cite{RateFuncEstimationAuxTargets}. 
Furthermore, this relationship between relaxation and fluctuation has been shown to hold for NESS Markovian processes in the hydrodynamic limit, supposing a corresponding large deviation principle exists ~\cite{HydrodynamicNoneqFluctDissTheory,MinimumDissPrinciple,MacroFlucTheory}. Indeed, the term $I_A(\rho)$ in the rate function Eq.~(\ref{eq:ratefunc}) can be identified as a non-adiabatic entropy production term vanishing upon relaxation to the steady state~\cite{NonAdEntropyMarkov,NonAdEntropyLangevin}.
Likewise, the rate functional can be related to an entropy flux and a quantity termed ``traffic''(which is a form of dynamical activity), both of which play a central role in the fluctuation-response theory for non-equilibrium diffusive systems~\cite{NonEqFluctuationsResponse1,NonEqFluctuationsResponse2}.
 Tentatively, from a NESS design point of view, all these observations suggest that a fluctuation level approach might be generally useful beyond exact knowledge of the corresponding dynamical rate function. 

\section{Conclusion}
\label{sec:Conclusion}
In this work, we leveraged dynamical fluctuations in an overdamped Brownian system to design a variety of non-trivial non-equilibrium steady-states (NESS). 
In particular, we envisioned the target state as a far-off fluctuation that is ``relaxed'' to a desirable steady-state by minimizing the known rate functional with respect to an adjustable force field parameters $\THETA$.
We demonstrated such approach successfully replicates the analytic solution of a particle in a harmonic potential driven by an angular field, 
and applied it to the design of non-trivial steady states in the form of prescribed occupation probabilities and steady-state fluxes tracing rose-curve and polygon motifs. We showed all targets could be successfully achieved, including the polygon motifs which were expected to be more difficult due to the presence of sharp turns and hard-edge linear flow.
Furthermore, satisfactory solution could be achieved even if the prescribed flux field was strictly unrealizable though still plausible for the motif constraint.  
Altogether, our method demonstrates that a dynamic fluctuation point of view provides a feasible entry into NESS design,
though much work is needed in extending it to useful scenarios such as interacting particle ensembles. 
For future studies, it will be interesting to understand how this design view may tie into more general frameworks of non-equilibrium fluctuation response or established entropy production relations.  

$\qquad$ 

\begin{acknowledgments}
The authors thank Daniel Nickelsen, Christian Maes, and Juzar Thingna for helpful discussions. This work was publicly funded through the Institute for Basic Science, South Korea, Project Code IBS-R020-D1.
\end{acknowledgments}

%
\end{document}


\renewcommand{\theequation}{S\arabic{equation}}
\renewcommand{\thetable}{S\arabic{table}}
\renewcommand{\thefigure}{S\arabic{figure}}
\renewcommand{\bibnumfmt}[1]{[S#1]}
\renewcommand{\citenumfont}[1]{S#1}

	\begin{center}
	\textbf{\large Supplementary Material} 
	\end{center}

\def\btheta{{\boldsymbol\theta}}
\def\st{{\mathrm{st}}}

\def\eg{{\it{e.g.,~}}}
\def\ie{{\it{i.e.,~}}}
\def\kB{{k_{\mathrm{B}}}}
\def\bj{\mathbf{j}}
\def\bx{\mathbf{x}}
\def\by{\mathbf{y}}
\def\T{\mathcal{T}}
\def\xdot{\mathbf{\dot{x}}}
\def\bjst{\bj_{\mathrm{st}}}
\def\pst{p_{\mathrm{st}}}
\def\Bf{{\mathbf{f}}}

\def\TT#1{\color[rgb]{0.00,0.18,0.65} [#1]\color{black}~}
\def\wp#1{\color[rgb]{0.18,0.0,0.88} [#1]\color{red}~}

\section{Simulation Details}
Optimization of the procedure outlined in the main text is carried out using an in-house code interface to a Brownian simulation. Gradient update step sizes $\alpha_{A/B}$, corresponding to $\nabla_{\btheta} I_{A/B}$ respectively, were taken as $\alpha_A \approx 0.04\mathrm{-}0.003$ with $\alpha_A > \alpha_B$ to emphasize probability term contributions. 

Following the expressions of $\nabla_{\btheta}I_{A/B}$ in Eqs. (8) and (9) in the main text, evaluation of the derivatives reduces to evaluation of $\nabla_{\btheta} \Bf$ and $\nabla_{\btheta}\nabla \ln p_{\st}(\bx|\btheta)$. 
The first can be done analytically following the chosen forms of $\Bf$ as elaborated in the next section, while the second is done directly in a Brownian simulation using a finite central difference approximation as
\begin{equation}
\frac{\partial g(\btheta)}{\partial \theta_i} \approx \frac{g(\cdots,\theta_i+h,\cdots) - g(\cdots,\theta_i-h,\cdots)}{2 h}~, 
\end{equation} 
where $g(\btheta) \equiv \nabla\ln p_{\st}(\bx|\btheta)$ and $h$ is the derivative step size. 
In general, values of $h=0.02\mathrm{-}0.05$ proved adequate for convergence and accuracy given the Brownian set up below. Parameter solutions are approximately converged within $N=2\mathrm{-}3\times10^3$ iterations, with longer iterations yielding more refined solutions. Exact parameters for each target are as follows. Roses: $\alpha_A=0.04$, $\alpha_B=0.02$, $h=0.03$, $N=6000$. Polygons: $\alpha_A=0.01$, $\alpha_B=0.003$, $N=3000$ with $h=[ 0.02,0.04,0.05 ]$ for triangle, hexagonal, and square respectively. 

Brownian simulations are carried out for an overdamped particle of mass $m=1$ with mobility $\chi=1$ in two dimensions. Dynamics are integrated using a standard Euler-Maruyama algorithm with time step $dt=0.005$ and $\beta=2$. For steady state averages and derivative evaluation during iterative optimization, $100$ independent trajectories consisting of $10^5$ time steps are used, with the exception of the triangle target that used $200$ independent trajectories. Final optimized steady state quantities in reported figures are found from $2000$ independent, $10^5$ time-step trajectories. For optimization, evaluation of integrals and histograms were carried out in inside a square area with limits set to $[-3,3]$ on both axis, and divided by a uniform grid size of $0.1$ units.

\section{Force Field Details}
	The model force field $\Bf$, where parameters are left implicit, is taken to be of the general form 
\begin{equation} 
\Bf = \Bf_h + \Bf_{an} + \Bf_o + \Bf_s
\end{equation} 
where $\Bf_h$ is a restoring harmonic force, $\Bf_o$ is a repulsive force at the origin, $\Bf_{an}$ is an angular driving force, and $\Bf_s$ are a series of oscillating sinusoidal forces whose exact expression is specific on the target motif. Evaluation of $\nabla_{\btheta} \Bf$ follow from standard partial derivatives on the expressions below.  

For the rose-motif targets, the first three $\Bf_i$ are defined as 
\begin{equation}
\Bf_h =\theta_1 r \mathbf{\hat{r}}, \qquad 
\Bf_{an} =\theta_2 r \boldsymbol{\hat{\phi}},  \qquad 
\Bf_o =\theta_3 \mathbf{\hat{r}}~,  \\
\label{eq:f}
\end{equation}
with the last term 
\begin{equation}
\Bf_s = \mathbf{A}_{a,b} \cdot\cos(\btheta') \mathbf{\hat{r}} + \mathbf{A}_{c,d} \cdot\cos(\btheta') \boldsymbol{\hat{\phi}}~,
\end{equation}
where 
\begin{equation}
\mathbf{A}_{i,j} \equiv (\btheta_i + \btheta_j/r)~,
\label{eq:Aij}
\end{equation} 
and 
\begin{equation}
\cos(\btheta') \equiv
\cos(\btheta_{I}r+\btheta_{II}\phi+\btheta_{III})~. 
\label{eq:cos}
\end{equation} 
For all expressions, hatted symbols denote polar coordinate unit vectors.
Here the dotted operation means an element-by-element product of two equal-sized parameter sets $\btheta_{\alpha}$ and $\btheta_{\beta}$,
e.g. $\btheta_{\alpha}\cdot\cos(\btheta_{\beta}) \equiv \sum_i^{n} \theta_{\alpha i} \cos(\theta_{\beta i})$, where $n$ represents the total number elements. 

Analogously, for the polygon motifs we have 
\begin{equation}
\begin{aligned}
\Bf_h &=  \theta_1 x~\mathbf{\hat{x}}  +\theta_2 y~\mathbf{\hat{y}}, \\
\Bf_{an} &= -\theta_3 y~\mathbf{\hat{x}}  +\theta_4 x~\mathbf{\hat{y}}, \\ 
\Bf_o &=  \theta_5 x/r~\mathbf{\hat{x}}+\theta_6 y/r~\mathbf{\hat{y}},
\end{aligned}
\end{equation}
and $\Bf_s$ in this case  
\begin{equation}
\Bf_s =  [\mathbf{A}_{a,b}\cdot\cos(\btheta')] x~ \mathbf{\hat{x}} + 
       [\mathbf{A}_{c,d}\cdot\cos(\btheta')] y~ \mathbf{\hat{y}}~,
\end{equation}
where $\mathbf{A}_{i,j}$ and $\cos(\btheta')$ are same as above in eq. \ref{eq:Aij} and eq. \ref{eq:cos} respectively. Hatted symbols now denote Cartesian unit vectors. We note that for the triangle motif we needed additional terms in $\Bf_s$ of the form 
\begin{equation}
\Bf_s^{\mathrm{add}} = -\btheta_e\cdot\cos(\btheta') y~\mathbf{\hat{x}} 
		   ~ +\btheta_f\cdot\cos(\btheta') x~\mathbf{\hat{y}}~,
\end{equation}
which essentially represent sinusoidal modifications to the driven angular field. 

Due to the large number of parameters, all parameters for all targets are presented below in figures where the $y$ axis represents the magnitude of parameter $\theta_i$. 
We note that, due to the inherent stochasticity of the method, parameter values will oscillate around $0.02$ units of its final value,
which corresponds to about the size of the marker in graphs below.

\subsection{Rose-motif Parameter Figures}
\begin{figure*}[htbp!]
\includegraphics[scale=0.455]{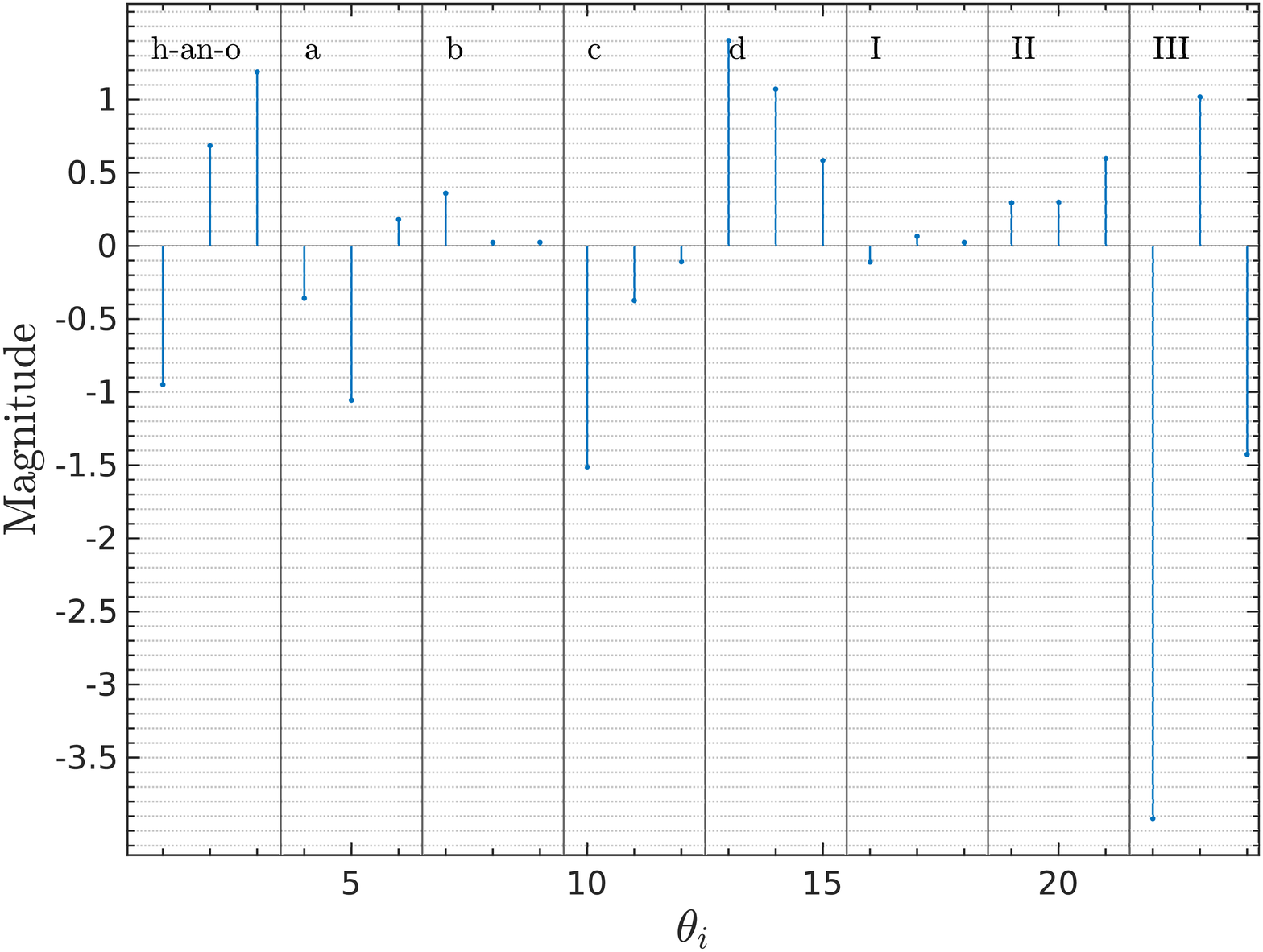} 
\caption{Parameter plots for rose-motif with $k=3$. Here, the x-axis corresponds to parameter $\theta_i$ number and y-axis its magnitude with a total of 27 parameters ($n=3$ sinusoids). Vertical black lines are added to demarcate the respective parameter set $\btheta_\alpha$ and the added text denotes the label of $\alpha=\{$a,b$,\cdots\}$ etc. The labels `h an o' denote the parameters for $\Bf_h$, $\Bf_{an}$ and $\Bf_o$ in eq. \ref{eq:f} in eq. \ref{eq:f}. Note parameters for $\btheta_{II}$ have been scaled down by a factor of $10$ in figure.} 
\label{fig:k3}
\end{figure*} 

\begin{figure*}[htbp!]
\includegraphics[scale=0.455]{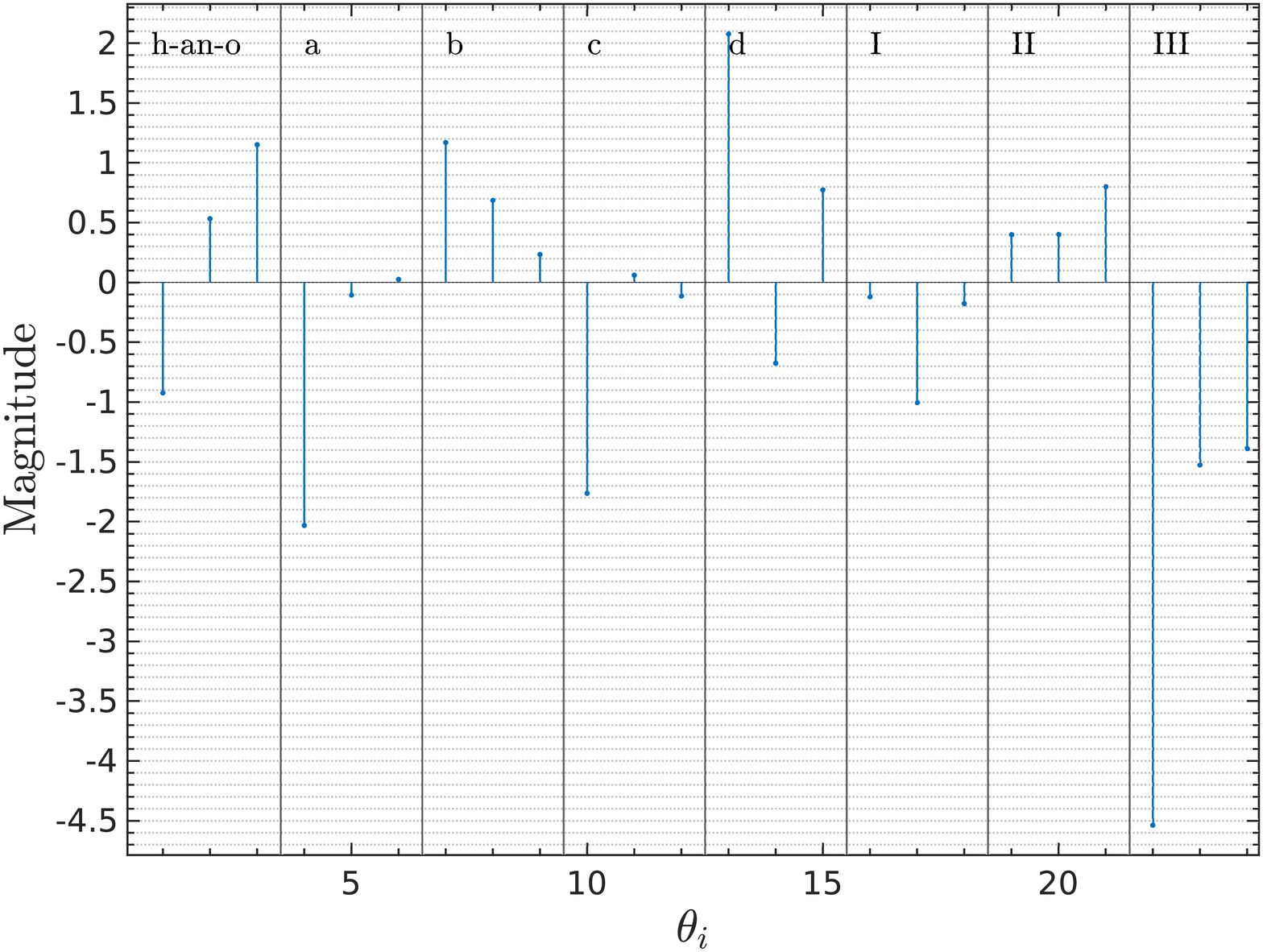} 
\caption{Parameter plot for rose-motif $k=4$. Description as in figure \ref{fig:k3}. }
\label{fig:k4}
\end{figure*} 

\begin{figure*}[htbp!]
\includegraphics[scale=0.455]{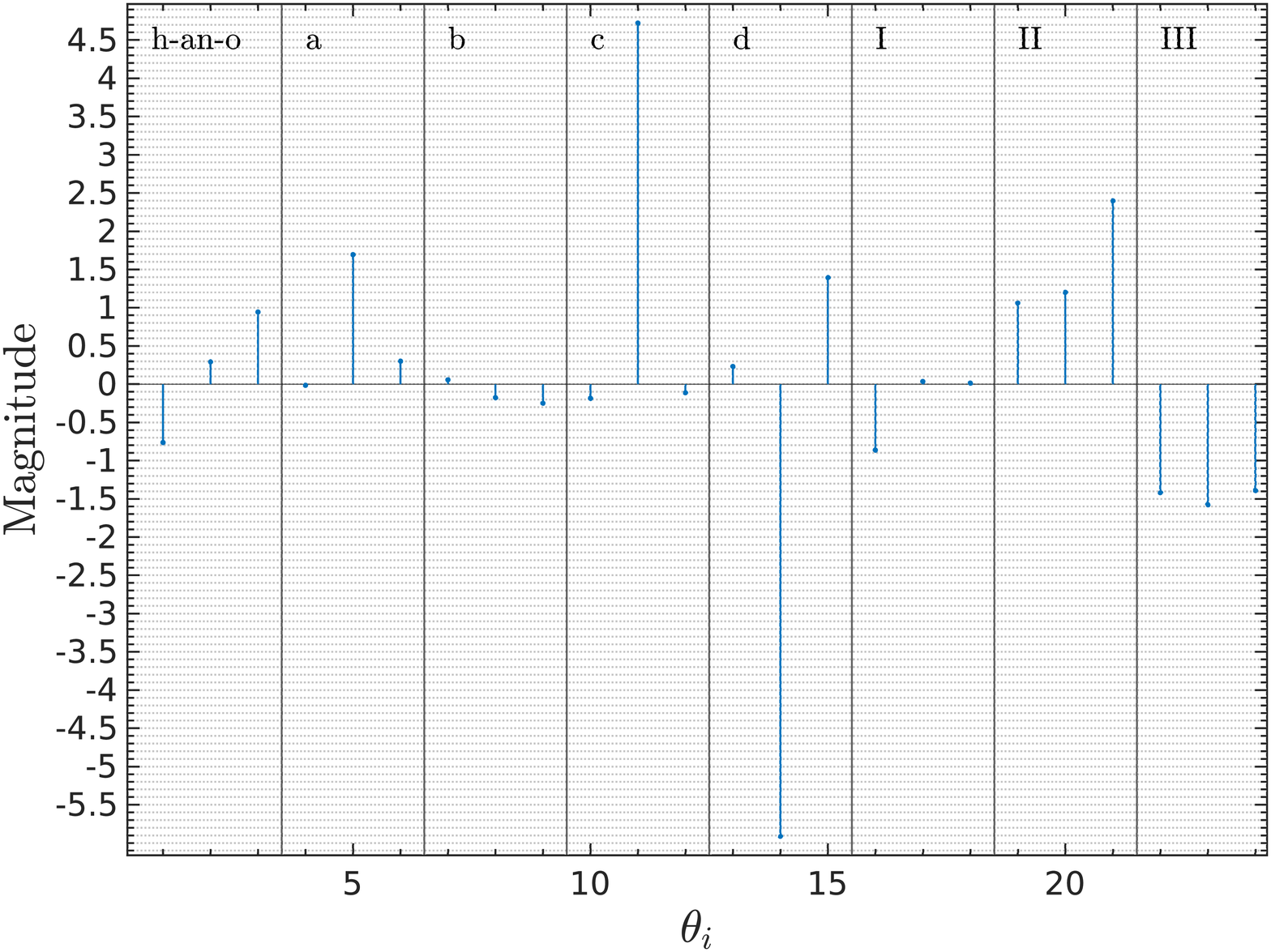} 
\caption{Parameter plot for rose-motif $k=12$.  Description as in figure \ref{fig:k3}.}
\label{fig:k12}
\end{figure*} 

\clearpage 

\subsection{Polygon-motif Parameter Figures}

\begin{figure*}[htbp!]
\includegraphics[scale=0.455]{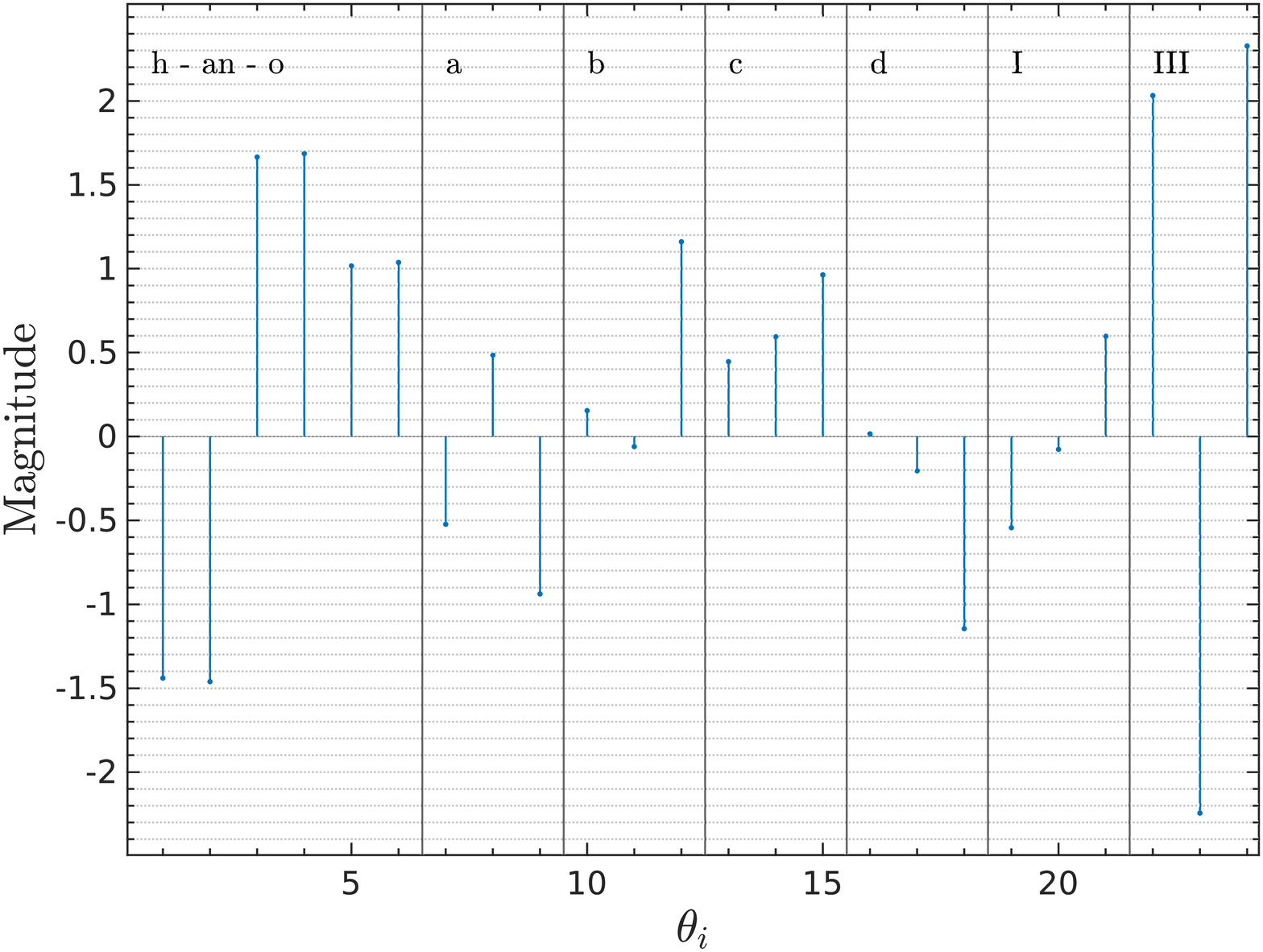} 
\caption{Parameter plots for the square motif. Here, x-axis corresponds to parameter $\theta_i$ number and the y-axis its magnitude, with a total of $24$ optimized parameters ($n=3$ sinusoids). Vertical black lines are added to demarcate the respective parameter set $\btheta_\alpha$ and added text denotes the label of $\alpha=\{$a,b$,\cdots\}$ etc. The labels `h an o' denote the parameters $\Bf_h$, $\Bf_{an}$ and $\Bf_o$ in eq. \ref{eq:f}.
 Parameters $\btheta_{II}$ are fixed and taken as $\btheta_{II}=\{2,4,6\}$.}
\label{fig:sq}
\end{figure*} 

\begin{figure*}[htbp!]
\includegraphics[scale=0.455]{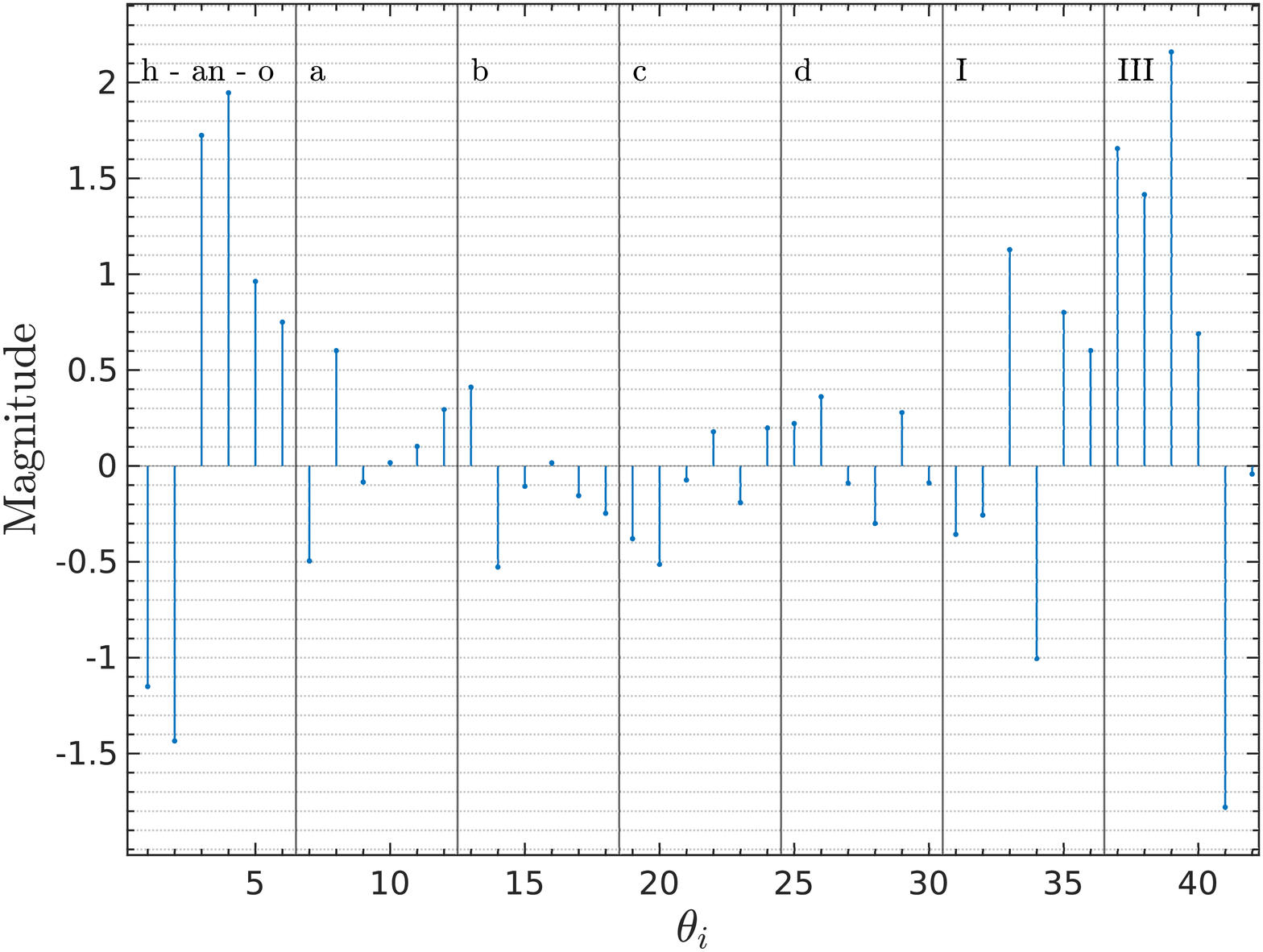} 
\caption{Parameter plots for hexagon motif with a total of $42$ optimized parameters ($n=6$ sinusoids). Here $\btheta_{II}$ are fixed and taken as $\btheta_{II}=\{2,4,6,8,10,12\}$.} 
\label{fig:hex}
\end{figure*} 

\begin{figure*}[htbp!]
\includegraphics[scale=0.365]{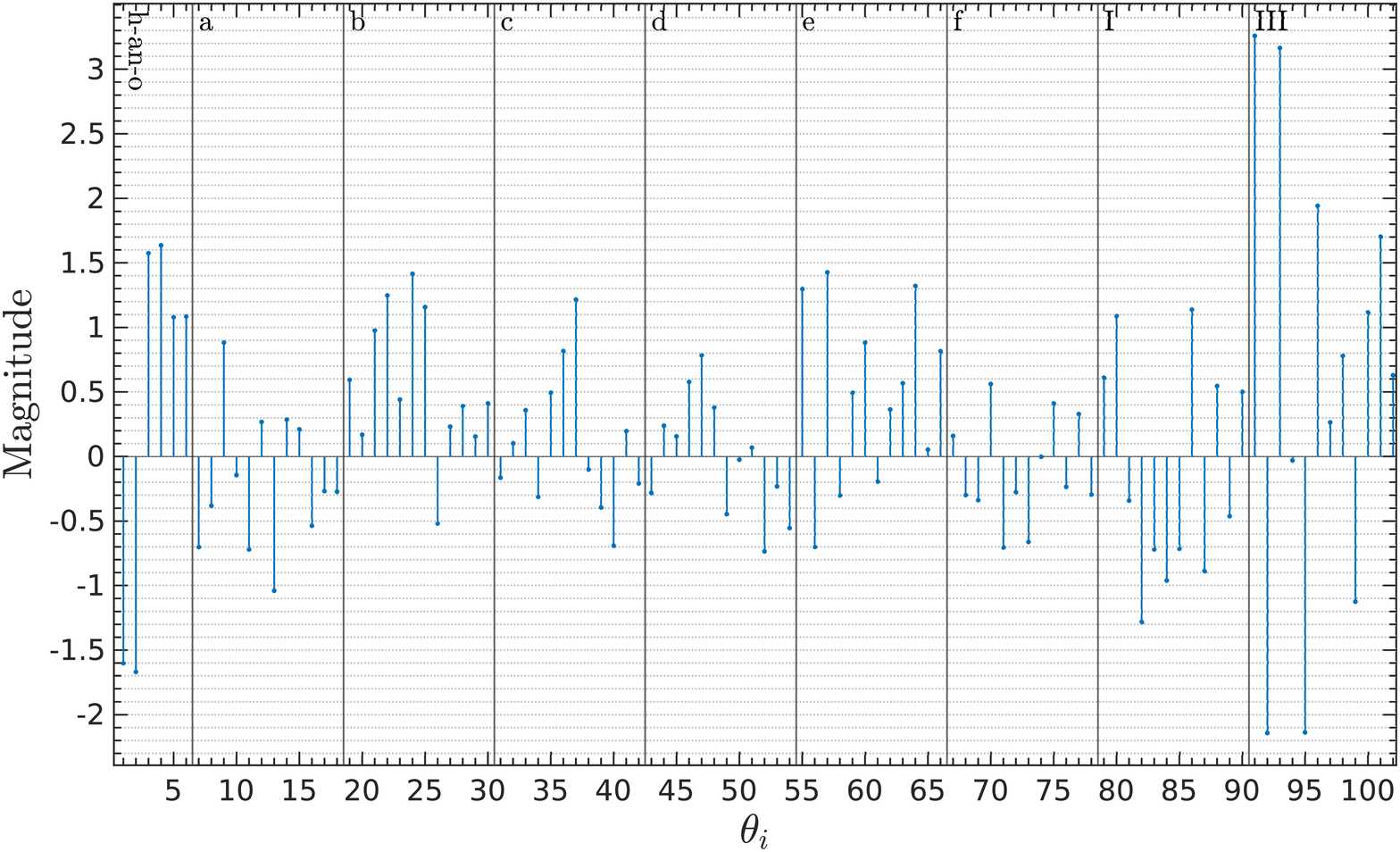} 
\caption{Parameter plots for the triangle motif with a total of $102$ optimized parameters ($n=12$ sinusoids). Here $\btheta_{II}$ are fixed and doubled as $\btheta_{II}=\{1,1,3,3,5,5,7,7,9,9,11,11\}$.}
\label{fig:sq}
\end{figure*} 
